# Transformation from 5G for Verticals Towards a 6G-enabled Internet of Verticals

Maziar Nekovee
Email: m.nekovee@sussex.ac.uk
6G Lab, School of Engineering and Informatics, University of Sussex UK

*Abstract*

5G will enable and greatly accelerate digital transformation of vertical sectors. In the longer term this will evolve from per-vertical connectivity services in 5G to the emergence of the 6G-enabled Internet of Verticals (6G-IoV). In this paper we describe and examine enabling technologies for this further transformation of verticals and examine some examples of 6G-IoV, next generation cloud manufacturing and manufacturing as a service, next generation smart energy grids and the Internet of Robotics (IoR).

I. INTRODUCTION

5G in conjunction with IoT, AI and Cloud technologies will drive a significant digital transformation to fulfill the requirements of the fully connected, digitalized carbon-neutral society and industries [1]. Unlike the previous generations, 5G is offering not only ultrafast (20 Gbps peak data-rates) but also ultra-reliable, ultra-low-latency (1ms) and massive connectivity capabilities. In comparison with previous generations, 5G also comes with significant architectural innovations including network slicing, private networks and edge computing [1]. Enabled by network slicing: in the 5G era communication services are transitioning from mass connectivity provisioning to per service level and vertical customization. Edge solutions and private networks help to overcome network performance limits, e.g. in term of latency and deterministic delivery. Local data processing enabled by 5G private networks help to achieve the stringent security and privacy requirements of vertical sectors.

In the longer term the evolution of 5G "networkiation" of verticals towards the 6G era, combined new capabilities provided by AI cloud and edge computing will pave the way for a new type of the Internet, which call the 6G-enabled Internet of Verticals (6G-IoV) . The 6G-IoV virtualizes, orchestrates and federates a multitude of vertical entities and services, hence ushering in entirely new, and previously unimaginable, services, businesses and actors, including new OTTs.

The rest of this paper is organized as follows. In Section II we review key enabling technologies for the vision 6G-IoV. We then illustrate 6G-IoV by describing a number of examples from manufacturing, This is followed by a description of examples of 6G-IoV from the fields of cloud manufacturing, smart grids and robotics in Section III. We conclude this paper in Section V.

II. 5G ENABLES FOR VERTICALS

*A. Network Slicing*

Network slicing [2,3] is an innovative architectural solution in 5G allowing operators to provide multiple services with different performance characteristics on the same physical infrastructure. Each network slice operates as an independent, virtualized version of the network designed to serve a defined business purpose or customer. GSMA [3] has identified the following vertical sectors as ones that will benefit from network slicing: Augmented Reality and Virtual Reality, Automotive, Energy, Healthcare Manufacturing, Internet of Things, Public Safety, Smart Cities.

*B. Private (Non-public) Networks*

In the shorter term, 5G private networks can fulfil many of the requirements of Network Slicing for verticals, as well as offering additional security and privacy advantage to such sectors. Private 5G networks comprise Radio Access Networks (RAN) and core elements. The 5G base station (gNB) can scale from low to high capacity and power output according to needs. They connect to a private core and edge networks (in contract to operator's public core) that provided security, authentication, session management and QoS-control. The private 5G core network could be deployed on edge compute nodes, installed locally on premises to ensure high reliability, enable low-latency, ultra-security, and privacy, or be virtualized and reside on cloud servers. In the longer term, 5G private networks may coexist with Network Slicing or they may beentirely overtaken by slicing.

*C. Edge Computing/Edge AI*

ETSI is standardizing Multi-access Edge Computing (MEC) [4]. The benefits are reduced latency, more efficient network operation for certain applications, and an improved user experience.

For many vertical applications, latency will determine how close the edge servers need to be to user devices. 5G networks are striving for 1 msec latency (round-trip time) within the network. Light travels 300 km per millisecond, so designers will need to plan their applications accordingly. A robotic controller, for example can tolerate around four milliseconds latency (round trip) and therefore, allowing for fluctuations and processing time, MEC need to be placed as close as possible to the controller. The other consideration is the amount of data that needs transportation to MEC . For example, a factory performing AI-based video analytics of its assembly line operation may wish to do such calculations on a MEC at the factories location rather than backhauling a huge amount of data to a more central operator location.

*D. URLLC and Deterministic Networking*

When a network can provide end-to-end ultra-reliable packet transmission with bounded small values of latency/jitter, it is said to be a deterministic network. One of the most important use cases of 5G Ultra Reliable Ultra Low Latency (URLLC) communication is in smart and fully automated manufacturing. As an example, consider mobile robots such as Automatic Guided Vehicle (AGV), which have numerous applications in the future, factories. They are usually monitored and controlled by a guidance control system. The mobile network is the most promising communication technology due to the large-scale mobility of the vehicles. The communication in some mobile robot applications may require the transmission latency to be between 1 to 10 msec and jitter to be less than 50 % of latency. Furthermore, the reliability is required to be above six nines (99,9999 %). For URLLC local area networking (e.g. inside a single factory plant) Time Sensitive Networking standardised in IEEE 802.1 [5] can be used. However, wide area deterministic networking for evolved vertical applications, including distributed cloud manufacturing and wide-area smart grids would require expanding the capabilities of TCP/IP [6,7].

III. SCENARIOS FOR THE INTERNET OF VERTICALS

A. *Next generation cloud manufacturing*

Cloud manufacturing(CMfg) [8] is an emerging manufacturing paradigm developed from existing advanced manufacturing models and enterprise information technologies with the support of cloud computing, service-oriented technologies, and advanced computing technologies. It transforms manufacturing resources and manufacturing capabilities into manufacturing services, which can be managed and operated in an intelligent and unified way to enable the full sharing and circulating of manufacturing resources and manufacturing capabilities. Cloud manufacturing can provide safe and reliable, high quality, cheap and on-demand manufacturing as a services. In CMfg system [11], various manufacturing resources and abilities can be intelligently sensed and connected through wide area networks , and automatically managed and controlled using 5G and AI technologies

.We envisage that, starting already with 5G and fully evolving in the 6G era, URLLC combined with wide area deterministic networking will enable distributed network architecture for cloud manufacturing and new Over The Top (OTT) players in the manufacturing sector (Mf-OTT) . This model builds in analogy with the how the internet architecture together with cloud computing has enabled the emergence of OTTs such as Uber for the transport sector. As shown in Fig. 1, it introduces three layers: manufacturing service providers, Mf-OTTs and end-users. A layer of manufacturing service providers consists of manufacturing, computational, and AI clouds. Manufacturing clouds are formed by 5G/beyond-5G interconnected manufacturing service providers . The Mf-OTT use intelligent search mechanism analogous to Internet search engines endowed with new AI capabilities for dynamic manufacturing service composition based on end-user requirements The search mechanism propagates manufacturing service inquiries through the decentralized network. The search mechanism distributes matching, scheduling, and rational decision-making procedures between service providers and end-users. Software and AI supporting the search process are accessible to Mf-OTTs providers in computation and AI clouds which are accessed through the 6G network.

B. *Next generation smart grids*

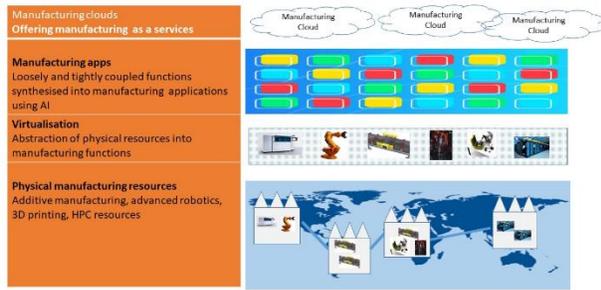

**Figure 1 Distributed cloud-manufacturing architecture enabled by 5G wide area networks.**

The growth in end-use consumption results in electricity generation increasing 79% between 2018 and 2050. Electricity use grows in the residential sector as rising population and standards of living in non-OECD countries increase the demand for appliances and personal equipment. Electricity use also increases in the industry and business sectors as well as seeing a steep rise in transportation sector as plug-in electric vehicles (EVs) enter the fleet and electricity use for rail expands. Consequently besides residential sector , industry and enterprise verticals can be expected to become a key component of energy prosumer by 2050 while EVs are expected to play a major role in both prosumer and energy storage for future smart grids. Due to their energy intensity, industrial prosumers are important players in the energy transition. Their existing assets often already have a significant amount of flexibility, and installing Distributed Energy Resources (DERs) onsite (e.g. renewables, storage) can further help their business case. In recent years, three has been an exponential growth in offshore wind power turbines as a key renewable energy technology and there is great interest recently in floating wind power turbines as they can b reconfigured based on wind speed, water depth and distance to shore [9].

The above evolving landscape will necessitate the availability of a unified, high stability and ultra-reliable 6G network, which also support high-bandwidth and low-latency requirements of next generation smart grids.

*C. The Internet of Robots*

Robotics is a rapidly growing technology with many new applications appearing in the market every day. According to a recent report from the International Federation of Robots (IFR), the largest projected market for Robotics currently is in the manufacturing sector, with the highest level of development being in automotive and electronic industries [10].

Robots endowed with Artificial Intelligence, rich sensorial capabilities and mobility will require communicating in a variety of scenarios. Robots may use various forms of multi-sensory/multi-modal communications which could include 3D images and videos , sound and ultra-sound, temperature and haptics to share situational information with other robots and humans. Or they may develop new forms of speech or multi-layer visions perception for their communications which are unrecognizable by humans.

Cloud robotics is a paradigm that leverages the powerful computation, storage and resources of modern data centres combined with high speed and low-latency communication to enhance the capabilities of robots. Cloud robots are controlled by a "brain" in the cloud that may constitute a data centre, a shared knowledge base, artificial intelligence and deep learning algorithms, information processing, task planners, environment models, etc. Cloud Robotics open up the possibility of *robot virtualization,* where all or the majority of Robotic intelligence as well as low-level control functions are virtualized and run on cloud or edge cloud servers. Table 1 summarises key robotic Communication scenarios together with their projected data-rates.

| Scenario | Purpose | Expected communication rate |
|---|---|---|
| Robot-environment Communications Robot-Things Communications | Control of and adaption to Environment /exchange of sensory data with environment | Mbps/Gbps |
| Robot-Human Communications | Control, Cooperation/Coordination, Information and context sharing, Problem solving | Up to Gbps |
| Robot-Robot Communications (voice, video, data, VR/AR, Holographic, Multi-modal/Multi-sensory l | Control, Cooperation/Coordination, Information and context, sharing, Problem solving | Up to Tbps |
| Robot-Cloud Communications | Control, Information Sharing, Virtualization, AI | Up to Tbps |

**Table 1. Robotic communication scenarios.**

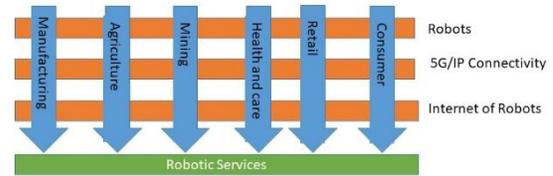

**Figure 2 The Internet of Robotics (IoR).**

The advancement of 6G networks to support and enable full cloudification, and virtualisation of robotic industry will pave the way for the transformation of robotic technology from specialized verticals to horizontal and virtualised robotic service platforms and ultimately an Internet of Robotics (IoR) , as illustrated in Fig. 2.

## IV. CONCLUSION

In this paper, we reviewed key architecture and networking innovations, which are underpinning the successful expansion of 5G connectivity services and platforms into vertical sectors and its advancement towards 6G. 5G for verticals is primarily aim at "5G native" connectivity services "per vertical". In the longer, we showed that this will pave the way for a transformation to the Internet of Verticals, and illustrate this concept with a few examples.

### REFERENCES

[1] 5G White Paper 2, NGMN Alliance, 27 July 2020, https://www.ngmn.org/work-programme/5g-white-paper-2.html, Accessed 29/11/2020.
[2] 3GPP, System architecture for the 5G System, Table 5.15.2.2-1, 3GPP TS 23.501 V16.4.0, 2020.
[3] GSMA, Network Slicing, Use Case Requirements, April 2018. https://www.gsma.com/futurenetworks/wp